\documentclass[letterpaper,conference]{IEEEtran}
\IEEEoverridecommandlockouts

\usepackage{cite}
\usepackage{xcolor}
\usepackage{url}
\usepackage{textcomp}
\usepackage{graphicx}
\usepackage{hyperref} 
\usepackage{multicol}
\hypersetup{
    linkbordercolor={0 0 1}
}
\makeatletter
\def\thickhline{%
  \noalign{\ifnum0=`}\fi\hrule \@height \thickarrayrulewidth \futurelet
   \reserved@a\@xthickhline}
\def\@xthickhline{\ifx\reserved@a\thickhline
               \vskip\doublerulesep
               \vskip-\thickarrayrulewidth
             \fi
      \ifnum0=`{\fi}}
\makeatother
\newlength{\thickarrayrulewidth}
\makeatletter
\def\bitthickhline{%
  \noalign{\ifnum0=`}\fi\hrule \@height \bitthickarrayrulewidth \futurelet
   \reserved@a\@xbitthickhline}
\def\@xbitthickhline{\ifx\reserved@a\bitthickhline
               \vskip\doublerulesep
               \vskip-\bitthickarrayrulewidth
             \fi
      \ifnum0=`{\fi}}
\makeatother

\newlength{\bitthickarrayrulewidth}
\usepackage{tablefootnote}

\usepackage{array}
\usepackage{multirow}
\usepackage{arydshln}

\usepackage{amsmath,amssymb,amsfonts}
\usepackage{stfloats}
\usepackage{nicefrac}

\usepackage{graphicx}
\usepackage{subfigure}
\usepackage{xcolor}
\usepackage{tikz,pgfplots}
\usetikzlibrary{arrows}

\usepackage[acronym]{glossaries}

\usepackage{algorithm}
\usepackage{algpseudocode}

\usepackage{caption}

\usepackage{amsmath}
\usepackage{mathtools}

\usepackage{enumitem}
\usepackage{array}

\usepackage{hyperref} 
\hypersetup{
    linkbordercolor={0 0 1}
}

\title{A Novel Transfer Learning-Based Approach for Screening Pre-existing Heart Diseases Using Synchronized ECG Signals and Heart Sounds\\
\thanks{\hspace{-1em}\rule{3cm}{0.5pt} \newline \textcopyright  \hspace{1pt} 2021 IEEE. Personal use of this material is permitted. Permission from IEEE must be obtained for all other uses, in any current or future media, including reprinting/republishing this material for advertising or promotional purposes, creating new collective works, for resale or redistribution to servers or lists, or reuse of any copyrighted component of this work in other works.}
} 
\author{Ramith Hettiarachchi$^{1}$, Udith Haputhanthri$^{1}$, Kithmini Herath$^{1}$, Hasindu Kariyawasam$^{1}$, Shehan Munasinghe$^{1}$, \\ Kithmin Wickramasinghe$^{1}$, Duminda Samarasinghe$^{2}$, Anjula De Silva$^{1}$ and Chamira U. S. Edussooriya$^{1}$ \\ \\ $^{1}$Department of Electronic and Telecommunication Engineering, University of Moratuwa, Sri Lanka \\
$^{2}$Lady Ridgeway Hospital for Children, Sri Lanka \vspace{-0.52cm}}

\begin{document}

\maketitle

\begin{abstract}
Diagnosing pre-existing heart diseases early in life is important as it helps prevent complications such as pulmonary hypertension, heart rhythm problems, blood clots, heart failure and sudden cardiac arrest. To identify such diseases, phonocardiogram (PCG) and electrocardiogram (ECG) waveforms convey important information. Therefore, effectively using these two modalities of data has the potential to improve the disease screening process. 
We evaluate this hypothesis on a subset of the PhysioNet Challenge 2016 Dataset which contains simultaneously acquired PCG and ECG recordings. Our novel Dual-convolutional neural network based approach uses transfer learning to tackle the problem of having limited amounts of simultaneous PCG and ECG data that is publicly available, while having the potential to adapt to larger datasets. In addition, we  introduce two main evaluation frameworks named \textit{\textbf{record-wise}} and \textit{\textbf{sample-wise}} evaluation which leads to a rich performance evaluation for the transfer learning approach. Comparisons with methods which used single or dual modality data show that our method can lead to better performance. Furthermore, our results show that individually collected ECG or PCG waveforms are able to provide transferable features which could effectively help to make use of a limited number of synchronized PCG and ECG waveforms and still achieve significant classification performance.

\end{abstract}



\vspace{-0.5cm}
\section{Introduction}
\label{sec:intro}


Pre-existing heart diseases are a range of abnormalities which affect the heart, that are either existing from birth (congenital heart diseases) or acquired later in life. The stethoscope is considered as the tool-of-choice ubiquitously used for initial medical examination of such complications by auscultation. However, accurate diagnosis using a stethoscope requires considerable expertise \cite{Mehmood2014}. 
In such occasions, an automated mechanism to help diagnose or screen for specific cardiac conditions can reduce the dependence on inadequate healthcare infrastructure and lead to significant improvements in the quality of patient care. While such a solution in general would deliver considerable healthcare benefits, the recent novel coronavirus pandemic, where persons with underlying conditions including those with cardiac complications are deemed to be at a higher risk of co-morbidity and mortality, has made this endeavor more critical\cite{Guo2020,Wang2020}.

To this end, 
studies have reported deep learning approaches for detecting cardiac abnormalities using either: (a) PCG signals \cite{Aziz2020, Dissanayake2020, Li2020} used for traditional stethoscope-based screening or, (b) ECG signals \cite{Huang2019} typically reserved for advanced diagnosis. 
Most deep learning techniques in the literature translate time domain PCG or ECG waveforms to higher dimensional representations depicting their time-frequency characteristics which are then fed to deep learning algorithms.
However, these studies do not evaluate the performance improvement that could be achieved via the integration of both PCG and ECG signals. This is especially important as there can be many occasions where the two types of signals convey mutually exclusive information regarding disease status \cite{SATO1966}. That is, a combination of ECG and PCG can provide accurate information about heart murmurs and other abnormal heart sounds that help to do accurate diagnosis. Therefore, integrating the analysis of PCG and ECG has the potential to significantly improve disease screening. 

\begin{figure*}[t!]
\begin{minipage}[b]{\linewidth}
\centering
\centerline{\includegraphics[width=0.92\columnwidth]{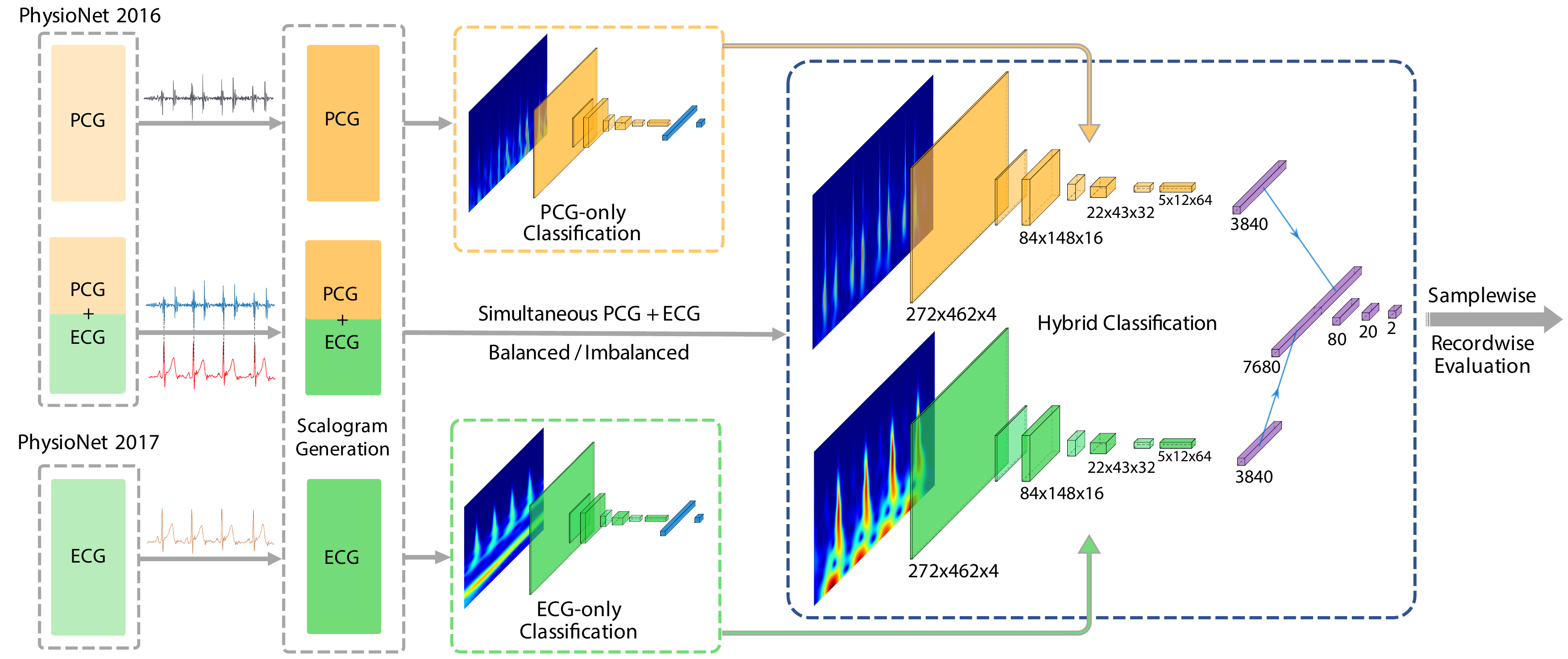}} 
\caption{\textbf{The proposed transfer learning based pipeline}: Balanced datasets created using setting-1 and setting-2 were fed into PCG-only and ECG-only models separately. The trained weights before the flatten layers of those models were used to initialize the feature extraction path weights of the hybrid model. The hybrid model was trained without freezing the feature extraction path weights using balanced or imbalanced datasets created based on setting-3. After obtaining the predictions from the hybrid model, sample-wise and record-wise evaluations were done separately.}
\label{fig:pipeline}
\vspace{-2ex}
\end{minipage}
\end{figure*}

A recent study by Chakir \textit{et al.} \cite{Chakir2020a} used traditional machine learning techniques to detect cardiac abnormalities from simultaneous PCG and ECG waveforms. By utilizing a subset of recordings from the PhysioNet \cite{Goldberger2000} Challenge 2016 (PHY16) \cite{Clifford2016} (training-a) dataset, they demonstrate that classification performance can be improved via integration of PCG and ECG beyond the sole use of PCG signals. 
However, usually traditional machine learning techniques have difficulty in performing well on large datasets\cite{Curtis2019}, whereas to tackle this problem through deep learning, there should be a sufficiently large set of simultaneous PCG and ECG recordings.  This creates a significant constraint for efficacious training and subsequent deployment: simultaneous PCG and ECG acquisition datasets are minimal, as such considerable effort and time has to be invested afresh to create new datasets for reliably training classifiers before its usage. Since devices capable of simultaneous PCG and ECG acquisition are either expensive or not yet widely adopted, this extra consideration can translate into years worth of new data collection before its deployment.  
 
To address this critical limitation, we propose a novel dual-convolutional neural network (CNN) based architecture which employs transfer learning \cite{Pratt1991}. This approach allows us to use large pre-existing datasets of individually acquired PCG and ECG waveforms, train two distinct CNNs to identify cardiac complications using each type of recordings, and transfer the learnt features onto a single integrated CNN. This can then be deployed for screening using dual-input PCG and ECG waveforms. We use the PHY16 (training-a) dataset for training, validating and testing our approach and demonstrate its feasibility for a reasonably eminent deployment.


\vspace{-0.1cm}
\section{Methods}
\label{sec:format}
Here, we present the proposed transfer learning based cardiovascular abnormality detection approach. Fig.~\ref{fig:pipeline} shows the overall pipeline where the generation of scalograms of the raw signals is followed by the hybrid classification model.





\subsection{Scalogram Generation}
\label{ssec:scalogram}
We employ scalograms to represent the time-frequency distribution of PCG and ECG signals. To this end, we apply a continuous wavelet transform (CWT) \cite{semmlow2014biosignal} to both ECG and PCG signals because it is suitable for analysing non-stationary signals using time-frequency representations. We select the Morlet wavelet \cite{web8} as the mother wavelet for the PCG signal transformation, since it has desirable properties of localization in both time and frequency domains of PCG signals \cite{Zin2003}. We choose the scaling parameter to be in the range of $7$-$130$, based on the frequency content of the the PCG signal that the wavelet should be sensitive to when applying the transform, to extract the frequency content of the signal in the time-frequency domain \cite{web8}. We select the complex Morlet wavelet \cite{web8} as the mother wavelet for the wavelet transform of the ECG signals. The complex wavelet allows both amplitude and phase information to be obtained in wavelet space and low values of the center frequency can be particularly useful for detecting short‐duration signal transients. \cite{Stiles2004} We set the bandwidth parameter and center frequency to $1.5$ and $1$ Hz, respectively. The range of the scaling parameter is set to $20$-$ 500$  (the reasoning was the same as for PCG signals). Example scalograms generated from the wavelet coefficients of a PCG and ECG signal are shown in Fig. \ref{fig:pipeline}. 





\subsection{Data Description}
\vspace{-0.05cm}
\label{ssec:datad}
We used the following three settings for creating datasets from the PHY16 and Physionet Challenge 2017 (PHY17) \cite{Clifford2017} datasets for model training.
\vspace{-0.03cm}
\begin{itemize}
    \item \hypertarget{d1}{\textbf{Setting 1}} - For this setting, we only included the PHY16 (training- b, c, d, e, f) datasets which contained only PCG recordings.
    We used a local peak detector \cite{wfdb} to detect the main peak in each heart cycle and non-overlapping samples of $3.5$ s windows were selected with respect to the detected peaks. Then, scalograms were generated for each of these samples.
    
    \item \hypertarget{d2}{\textbf{Setting 2}} - PHY17 dataset was used for this setting which contained only ECG recordings. 
    A QRS detector  \cite{Engelse1979} was used to detect R peaks \cite{Engelse1979} of each ECG signal. From the detected $n$ number of peaks, we then selected samples of $3.3$ s windows centering the $\left\lfloor \frac{n}{2} \right\rfloor$\textsuperscript{th}
    R peak. Finally, we generated scalograms for each selected sample of a particular record.
    
    \label{ssss:setting3}
    \item \hypertarget{d3}{\textbf{Setting 3}} - We used simultaneously recorded PCG and ECG signals available in the PHY16 (training-a) dataset ($407$ recordings) in this setting. Since the number of recordings were limited, the recordings were split into samples of $3.5$ s windows using the same approach employed in Setting 1. ECG signals were also segmented from the same points that the PCG signals were segmented. This effectively produced $3496$ simultaneous PCG and ECG samples. Finally, we generated scalograms from the segmented samples.
    
\end{itemize}
\vspace{-0.04cm}
The scalograms in each of the above settings were divided into sets of 70\%, 10\%, 20\% for training, validation and testing purposes respectively. This segregation was done randomly based on the record identities instead of sample identities. Therefore, we were able to make sure that the samples generated from the same record were not shared across the training, validation and test sets. Moreover, we also created imbalanced and balanced datasets for all the above settings.

 
 \vspace{-0.05cm}
 \subsection{Novel CNN Architecture} \label{ssec:novel_architecture}
 \vspace{-0.01cm}
 
The following CNN architectures were designed to classify PCG and ECG signals as {\it abnormal} and {\it normal}.

\label{ssec:subhead4}
 \subsubsection{PCG/ECG Classification}
 \label{ssec:pcg_ecg_models}
We implemented separate ECG-only, PCG-only classification models to classify abnormal/normal conditions of ECG, PCG datasets separately. Here, three repetitive downsampling blocks were implemented as feature extractors  followed by a convolutional layer along with a flatten layer. The downsampling block contained a stack of convolution, maxpooling and rectified linear unit (ReLU) layers each. The resultant feature vector was fed into a multilayer perceptron consisting of 3 dense layers followed by a softmax output.

 \subsubsection{Hybrid Model}
The intuition behind this hybrid model is to use a superimposed feature representation of both PCG and ECG signals. Therefore, this architecture comprises of separate PCG and ECG-feature extraction paths. To construct the hybrid model, outputs of the flatten layers of the two architectures used in ECG-only and PCG-only classification (subsection \ref{ssec:pcg_ecg_models}) were concatenated and then followed by a shared multilayer perceptron (See Fig. \ref{fig:pipeline}). 

\vspace{-0.05cm}
\subsection{Transfer Learning}
 \vspace{-0.01cm}
Since the number of simultaneous PCG and ECG recordings obtained from the PHY16 (training-a) data is too low (407 recordings) to train the hybrid model, we have adopted a transfer learning approach \cite{Pratt1991}. Initially, individual PCG-only and ECG-only models were trained on Setting 1 and Setting 2, respectively for both balanced and imbalanced datasets. To transfer the learnt knowledge, the weights before the flatten layers of both ECG-only, PCG-only models were then used in the PCG-feature extraction and ECG feature extraction paths of the hybrid model architecture (See Fig. \ref{fig:pipeline}).



\vspace{-0.05cm}
\subsection{Determining an Optimal Threshold for Classification}
\vspace{-0.01cm}

Obtaining predictions using the default threshold (0.5) on the softmax output of the model will not be the best method for imbalanced datasets \cite{Zhou2006}. Therefore we computed the optimal threshold such that it maximizes the G-mean metric given in (\ref{gmean}) for the validation set\cite{Kubat1997}. This effectively improved sensitivity and specificity parameters for the validation set which is a key requirement in medical diagnostic tests.
\begin{equation}
G\text{-}Mean\ =\ \sqrt{Sensitivity\times Specificity}
\label{gmean}
\end{equation}
The computed optimal threshold was then used to evaluate the model on the test set. Results showed that this method significantly removed the correlation between the imbalanced nature of the dataset and model performance (See Table \ref{tab:1}).




\setlength{\doublerulesep}{1\arrayrulewidth}
\begin{table*}
\centering
\caption{Comparison of previous approaches  vs. our experiments using standard
statistical evaluation parameters.}
\vspace{-0.19cm}
\label{tab:1}
\renewcommand{\arraystretch}{1.2}
\begin{tabular}{lllllllll} 
\hline
Author & Method & Database & \begin{tabular}[c]{@{}l@{}}Transfer\\Learning\end{tabular} & \begin{tabular}[c]{@{}l@{}}Optimal \\Threshold\end{tabular} & Input & \multicolumn{3}{c}{Results (\%)} \\ 
\hline\hline
\begin{tabular}[c]{@{}l@{}} Chakir \textit{et al.}\cite{Chakir2020a} \\ (2020)\end{tabular} & \begin{tabular}[c]{@{}l@{}}Support Vector Machine\\(SVM)\end{tabular} & \begin{tabular}[c]{@{}l@{}} PhysioNet 2016 (a)\\subset of 100 records\end{tabular} & - & - & \begin{tabular}[c]{@{}l@{}}record-wise\\(PCG+ECG)\end{tabular} & \begin{tabular}[c]{@{}l@{}}Sen = 92.31\\ Spe = 92.86 \end{tabular} & \begin{tabular}[c]{@{}l@{}}Acc ~= 92.5\\AUC = 95.05\end{tabular} & \begin{tabular}[c]{@{}l@{}}G-mean\\= 92.58$^{\star}$ \end{tabular}  \\ 
\hline
\begin{tabular}[c]{@{}l@{}}Li \textit{et al.}\cite{Li2020}\\(2020)\end{tabular} & CNN & \begin{tabular}[c]{@{}l@{}}PhysioNet 2016\\(a,b,c,d,e,f)\end{tabular} & - & - & \begin{tabular}[c]{@{}l@{}}record-wise\\(PCG only)\end{tabular} & \begin{tabular}[c]{@{}l@{}}Sen = 87\\Spe = 86.6\end{tabular} & Acc = 86.8 & \begin{tabular}[c]{@{}l@{}}G-mean\\= 86.8 \end{tabular}  \\ 
\hline 

\begin{tabular}[c]{@{}l@{}} Ren \textit{et al.}\cite{Ren2018} \\ (2018) \end{tabular} & \begin{tabular}[c]{@{}l@{}}Learnt VGG Net\\com/w a SVM\end{tabular} & \begin{tabular}[c]{@{}l@{}} PhysioNet 2016 \\ (PCG signals only) \end{tabular} & Yes & - &  & \begin{tabular}[c]{@{}l@{}}Sen = 24.6\\ Spe = 87.8 \end{tabular} & \begin{tabular}[c]{@{}l@{}} Acc = 56.2 \\\end{tabular} &
\begin{tabular}[c]{@{}l@{}}G-mean\\= 46.47 \end{tabular}  \\
\hline
\hline
\begin{tabular}[c]{@{}l@{}}Our Method [A]~\\\ \end{tabular} & Hybrid CNN & \begin{tabular}[c]{@{}l@{}}Setting 3\\(Imbalanced)\end{tabular} & Yes & Yes & \begin{tabular}[c]{@{}l@{}}record-wise\\(PCG+ECG)\end{tabular} & \begin{tabular}[c]{@{}l@{}}Sen = 87.72~\\ Spe = {\bf 87.5}\end{tabular} & \begin{tabular}[c]{@{}l@{}}Acc ~= 87.67\\AUC = {\bf 93.75} \end{tabular} & \begin{tabular}[c]{@{}l@{}}G-mean\\= {\bf87.6} \end{tabular} \\ 
\hline

\begin{tabular}[c]{@{}l@{}}Our Method [B]\\\ \end{tabular} & Hybrid CNN & \begin{tabular}[c]{@{}l@{}}Setting 3\\(balanced) \end{tabular} & Yes & \begin{tabular}[c]{@{}l@{}}Default\\thresh=0.5 \end{tabular} & \begin{tabular}[c]{@{}l@{}}record-wise\\(PCG+ECG)\end{tabular} & \begin{tabular}[c]{@{}l@{}}Sen = 85.7\\Spe = 82.6 \end{tabular} & \begin{tabular}[c]{@{}l@{}}Acc ~= 84.1\\AUC = 87.06\end{tabular} &  \begin{tabular}[c]{@{}l@{}}G-mean\\= 84.15 \end{tabular} \\
\hline

\begin{tabular}[c]{@{}l@{}}Our Method [C]\\\ \end{tabular} & Hybrid CNN & \begin{tabular}[c]{@{}l@{}}Setting 3\\(balanced)\end{tabular} & Yes &  {Yes} & \begin{tabular}[c]{@{}l@{}}sample-wise\\(PCG+ECG)\end{tabular} & \begin{tabular}[c]{@{}l@{}}Sen = 81.71\\Spe = 81.22\end{tabular} & \begin{tabular}[c]{@{}l@{}}Acc ~= 81.45\\AUC = 85.83\end{tabular} & \begin{tabular}[c]{@{}l@{}}G-mean\\= 81.47 \end{tabular} \\
\bitthickhline

\begin{tabular}[c]{@{}l@{}}Our Method [D] \\\ \end{tabular} & Hybrid CNN & \begin{tabular}[c]{@{}l@{}}Setting 3~\\(Imbalanced)\end{tabular} & No & Yes & \begin{tabular}[c]{@{}l@{}}record-wise\\(PCG+ECG)\end{tabular} & \begin{tabular}[c]{@{}l@{}}Sen = {\bf94.74}\\ Spe = 75\end{tabular} & \begin{tabular}[c]{@{}l@{}}Acc ~= {\bf90.41}~\\AUC = 91.06\end{tabular} & \begin{tabular}[c]{@{}l@{}}G-mean\\= 84.29 \end{tabular} \\ 

\hline
\begin{tabular}[c]{@{}l@{}}Our Method [E]\\\ \end{tabular} & Hybrid CNN & \begin{tabular}[c]{@{}l@{}}Setting 3\\(balanced) \end{tabular} & No & \begin{tabular}[c]{@{}l@{}}Default\\thresh=0.5 \end{tabular} & \begin{tabular}[c]{@{}l@{}}record-wise\\(PCG+ECG)\end{tabular} & \begin{tabular}[c]{@{}l@{}}Sen = 80.95\\Spe = 78.26 \end{tabular} & \begin{tabular}[c]{@{}l@{}}Acc ~= 79.55\\AUC = 82.5\end{tabular} & \begin{tabular}[c]{@{}l@{}}G-mean\\= 79.6 \end{tabular} \\ 
\hline
\begin{tabular}[c]{@{}l@{}}Our Method [F]\\\ \end{tabular} & Hybrid CNN & \begin{tabular}[c]{@{}l@{}}Setting 3\\(balanced) \end{tabular} & No & {Yes} & \begin{tabular}[c]{@{}l@{}}sample-wise\\(PCG+ECG)\end{tabular} & \begin{tabular}[c]{@{}l@{}}Sen = 78.86\\Spe = 77.67\end{tabular} & \begin{tabular}[c]{@{}l@{}}Acc ~= 78.23\\AUC = 86.04\end{tabular} & \begin{tabular}[c]{@{}l@{}}G-mean\\= 78.26 \end{tabular} \\ 
\thickhline
\begin{tabular}[c]{@{}l@{}}Our Method [G]\\\ \end{tabular} & PCG-only & \begin{tabular}[c]{@{}l@{}}Setting 3\\(balanced) \end{tabular} & No & \begin{tabular}[c]{@{}l@{}}Default\\thresh=0.5 \end{tabular} & \begin{tabular}[c]{@{}l@{}}sample-wise\\PCG\end{tabular} & \begin{tabular}[c]{@{}l@{}}Sen = 56.02\\Spe = 49.50 \end{tabular} & \begin{tabular}[c]{@{}l@{}}Acc ~= 52.69\\AUC = 55.26 \end{tabular} & \begin{tabular}[c]{@{}l@{}}G-mean\\= 52.65 \end{tabular} \\
\hline
\begin{tabular}[c]{@{}l@{}}Our Method [H] \\\ \end{tabular} & ECG-only & \begin{tabular}[c]{@{}l@{}}Setting 3\\(balanced) \end{tabular} & No & \begin{tabular}[c]{@{}l@{}}Default\\thresh=0.5 \end{tabular} & \begin{tabular}[c]{@{}l@{}}sample-wise\\ECG\end{tabular} & \begin{tabular}[c]{@{}l@{}}Sen = 70.16\\Spe = 81.73 \end{tabular} & \begin{tabular}[c]{@{}l@{}}Acc ~= 76.03\\AUC = 82.16\end{tabular} & \begin{tabular}[c]{@{}l@{}}G-mean\\= 75.72\end{tabular} \\ 
\hline

\multicolumn{9}{l}{$^{\star}$Distribution of class proportions is unknown}

\vspace{-0.4cm}
\end{tabular}
\end{table*}

\vspace{-0.06cm}
\section{Experiments} 
\vspace{-0.04cm}
\label{sec: experiments}






 \vspace{-0.01cm}
\subsection{Evaluation Criteria }
 \vspace{-0.01cm}

We conducted a comprehensive analysis to measure the robustness of our method based on the factors: 1. transfer learning, 2. balanced/imbalanced dataset setting, 3. predictions using optimal/default thresholding and 4. sample-wise/record-wise evaluation.

In sample-wise evaluation, final prediction scores were obtained based on the output of the hybrid CNN while for record-wise evaluation, the final prediction scores were obtained by aggregating the sample-wise predictions of samples which corresponded to a particular record.


\vspace{-0.1cm}
\subsection{Evaluation Metrics}
 \vspace{-0.02cm}

To measure the robustness of our method, we evaluated our models' performance on the test set with respect to Accuracy (Acc), Sensitivity (Sen), Specificity (Spe) \cite{Trevethan2017}, G-mean and Area under the Receiver Operating Characteristic Curve (AUC). We have focused on improving the G-mean score since it ensures that both sensitivity and specificity are higher (\ref{gmean}). Furthermore, based on the results, we have found that the G-mean score is practically a better measurement to capture the cases where the sensitivity is higher with a comparable specificity as well (Table \ref{tab:1}).


We then explored the effect of the above-mentioned factors using different training settings which included different model architectures and hyper-parameters.
The default training settings included: ADAM optimizer with a learning rate of 0.001, batch size of 20, categorical cross-entropy as the loss function with the architectures presented in section \ref{ssec:novel_architecture}. The training settings of the best results are described in section \ref{sec:results}.

 \vspace{-0.04cm}
\section{Results}
 \vspace{-0.02cm}
\label{sec:results}

Table~\ref{tab:1} shows a comparison based on different evaluation criteria explained in the section \ref{sec: experiments}. Our method [D] was able to achieve a high sensitivity of {\bf 94.74\%} while having a reasonable specificity of 75\% when using dilated convolution and dropouts. This was on par with the results of Chakir \textit{et al.} \cite{Chakir2020a} which used 100 simultaneous PCG and ECG records where the class distribution was unknown. Even though their selected handcrafted features were able to produce good results on this data, for a larger population, those features might not be able to capture properties of diverse abnormalities\cite{Xie2017}. Therefore, it poses a limitation to extend their method to large datasets. On the other hand, since the feature extraction is learnt by the CNN itself, it has the potential to capture properties of complex abnormalities. Furthermore, support vector machine's algorithm complexity hinders it from being applied to large datasets \cite{Cervantes2020}. 

The reason for the decreased specificity in method [D] is mainly due to class imbalance (117 Normal, 290 Abnormal) present in the dataset. However, our method [A] showed that even with an imbalanced dataset, it is possible to keep a fine balance between sensitivity and specificity by leveraging transfer learning. Performance improvements from methods [F] to [C] and [E] to [B] further strengthens this claim. 


Our method [A] outperforms the work done by Li \textit{et al.} \cite{Li2020} which uses only PCG data. However, a direct comparison is not possible due to different utilization settings of the PHY16 dataset. Our model used PCG and ECG feature extraction path weights obtained from best PCG-only {\it(PCG-22)} and ECG-only\textit{ (ECG-4)} models trained on setting 1 and 2, respectively. 
Furthermore, we outperform the work by Ren \textit{et al.}\cite{Ren2018} who followed a transfer learning approach on a VGG architecture.


Methods [B] and [E] which used the default threshold (0.5) were the best results obtained for the balanced record-wise setting. Meanwhile, methods [C] and [F] which used the optimal threshold were the best results for the balanced sample-wise setting. Even though the optimal threshold improved results significantly in the imbalanced dataset setting, in the balanced dataset setting it has not consistently given the best results.

Methods [A], [B], [D] and [E] show that record-wise evaluation gives better performance over methods [C] and [F] which uses sample-wise evaluation. This is because, the utilization of the complete time domain signals in record-wise evaluation makes it more robust to temporal distortions that may occur during the acquisition of signals. Methods [G] and [H] shows the results obtained from the models \textit{PCG-22} and \textit{ECG-4} when they were trained on setting 3 balanced dataset.

\begin{figure}[t!]
\begin{minipage}[b]{1.0\linewidth}
 \centering
 \centerline{\includegraphics[width=0.88\columnwidth]{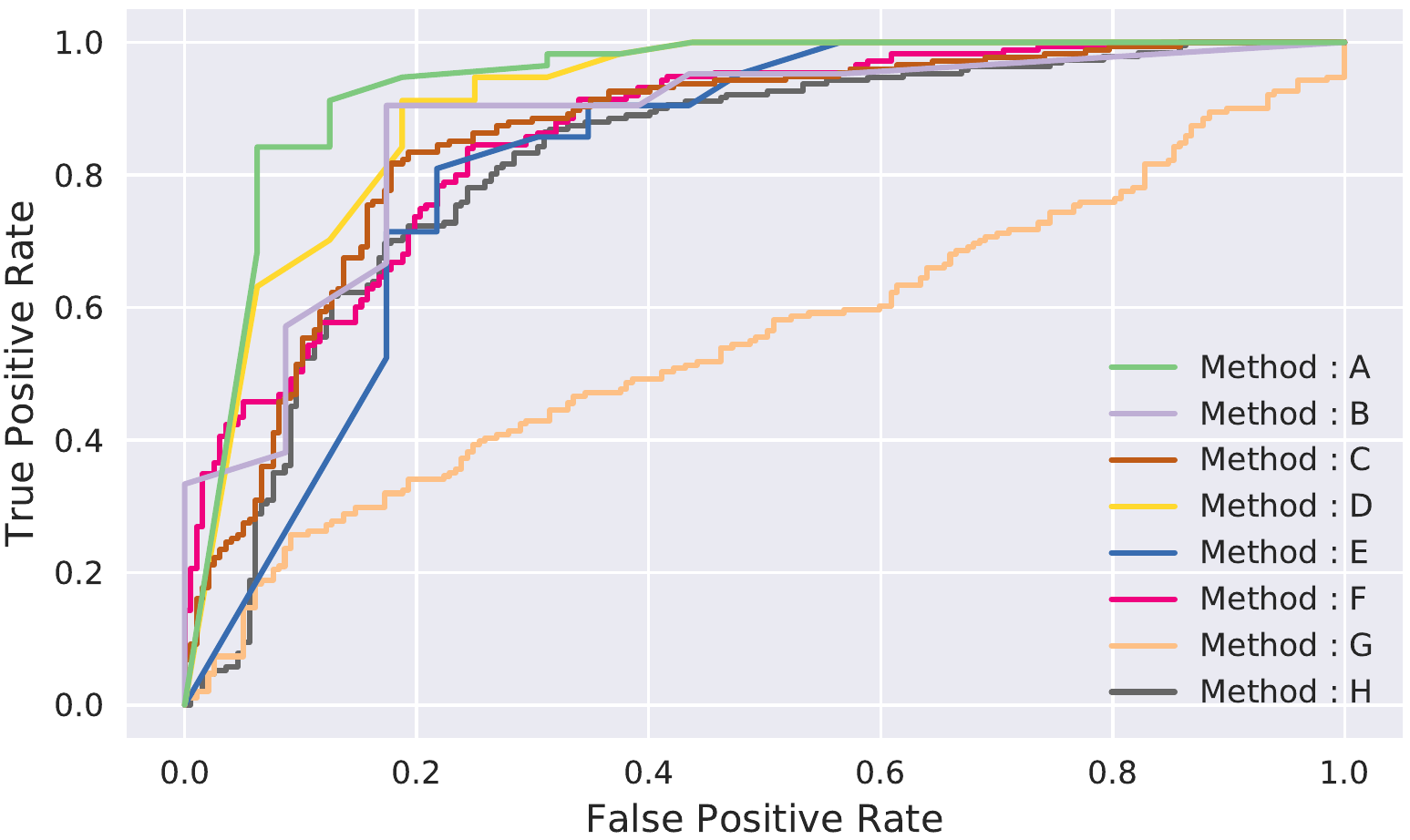}} 
\end{minipage}
\caption{Receiver operating characteristic curves }
\label{fig:ROC}
\vspace{-0.5cm}
\end{figure}

It can be clearly seen that the performances of all hybrid models are significantly better than the performances of PCG-only, ECG-only models. The AUC scores obtained from the receiver operating characteristic curves in Fig.~\ref{fig:ROC} have been summarized in Table~\ref{tab:1}, which shows that our methods achieved a maximum sensitivity of $94.74\%$ and a maximum G-mean score of $87.6\%$ with an AUC of $93.75\%$. 

\vspace{-0.2cm}
\section{Conclusion}
\vspace{-0.2cm}
We have demonstrated a proof-of-concept implementation of a novel classification framework for automated screening of cardiac complications by using dual-input PCG and ECG data. By employing transfer learning techniques, we circumvented the need to train our dual-input CNN solely on simultaneous PCG and ECG data, opening the potential to use vast amounts of individually acquired PCG and ECG waveforms contained in pre-existing datasets. Our current performance indices which were achieved only using 407 simultaneous PCG and ECG recordings, already show performance on-par to existing literature. We believe that this performance can be improved via expanded training with other well known online datasets which tabulate individual PCG or ECG data. 


\section{Acknowledgements}

Authors thank the University of Moratuwa for the financial support. Furthermore, authors thank Mr. Jathushan Rajasegaran for helpful suggestions and Mr. Vidura Dhananjaya for providing computing resources on Amazon AWS.


\IEEEtriggeratref{11} 
\bibliographystyle{IEEEbib}
\bibliography{strings,refs,IntelliscopeAI-Research}

\end{document}